\documentclass[showpacs,prd,preprint,nofootinbib,showkeys,unsortedaddress]{revtex4}
\usepackage{bm}
\usepackage{amsmath}
\usepackage{graphicx}
\usepackage{subfigure}
\usepackage[colorlinks=true,linktocpage=true,linkcolor=blue,citecolor=blue]{hyperref}
\usepackage[usenames,dvipsnames]{color}

\pdfoutput=1
\def\be{\begin{equation}}
\def\ee{\end{equation}}
\def\ba{\begin{eqnarray}}
\def\ea{\end{eqnarray}}
\definecolor{darkblue}{RGB}{0,0,160}

\newcommand{\checked}[1]{}

\begin{document}

\title{Shear-bulk coupling in nonconformal hydrodynamics}
\author{Gabriel S. Denicol}
\affiliation{Department of Physics, McGill University, 3600 University Street, Montreal,
QC H3A 2T8, Canada}
\author{Wojciech Florkowski}
\affiliation{Institute of Physics, Jan Kochanowski University, PL-25406 Kielce, Poland}
\affiliation{The H. Niewodnicza\'nski Institute of Nuclear Physics, Polish Academy of
Sciences, PL-31342 Krak\'ow, Poland}
\author{Radoslaw Ryblewski}
\affiliation{Department of Physics, Kent State University, Kent, OH 44242 United States}
\affiliation{The H. Niewodnicza\'nski Institute of Nuclear Physics, Polish Academy of
Sciences, PL-31342 Krak\'ow, Poland}
\author{Michael Strickland}
\affiliation{Department of Physics, Kent State University, Kent, OH 44242 United States}

\begin{abstract}
We compute the temporal evolution of the pressure anisotropy and bulk
pressure of a massive gas using second-order viscous hydrodynamics and
anisotropic hydrodynamics. We then compare our results with an exact
solution of the Boltzmann equation for a massive gas in the relaxation time
approximation. We demonstrate that, within second-order viscous
hydrodynamics, the inclusion of the full set of kinetic coefficients,
particularly the shear-bulk couplings, is necessary to properly describe the
time evolution of the bulk pressure. We also compare the results of
second-order hydrodynamics with those obtained using the anisotropic
hydrodynamics approach. We find that anisotropic hydrodynamics and
second-order viscous hydrodynamics including the shear-bulk couplings are
both able to reproduce the exact evolution with comparable accuracy.
\end{abstract}

\date{\today}
\maketitle


\section{Introduction}

\label{sect:intro} 

Dissipative hydrodynamics plays a central role in the phenomenology of the
quark gluon plasma. Since quantum mechanics implies that there is a lower
bound on the shear viscosity to entropy density ratio \cite%
{Danielewicz:1984ww,Kovtun:2004de}, one must include dissipative viscous
corrections in order to realistically model the spatiotemporal evolution of
the soft degrees of freedom of the system. The application of ideal \cite%
{Huovinen:2001cy,Hirano:2002ds,Kolb:2003dz} and second-order viscous
hydrodynamics \cite%
{Muronga:2001zk,Muronga:2003ta,Muronga:2004sf,Heinz:2005bw,Baier:2006um,Romatschke:2007mq,Baier:2007ix,Dusling:2007gi,Luzum:2008cw,Song:2008hj,Heinz:2009xj,El:2009vj,PeraltaRamos:2009kg,PeraltaRamos:2010je,Denicol:2010tr,Denicol:2010xn,Schenke:2010rr,Schenke:2011tv,Bozek:2011wa,Niemi:2011ix,Niemi:2012ry,Bozek:2012qs,Denicol:2012cn,Denicol:2012es,PeraltaRamos:2012xk,Calzetta:2014hra,Denicol:2014vaa}
now has a long history with recent developments focusing on constructing
complete and self-consistent methods for deriving the fluid-dynamical
equations of motion and the associated transport coefficients. Following a
different strategy, another promising framework for describing the soft
dynamics of relativistic systems has recently been developed called
anisotropic hydrodynamics~\cite%
{Florkowski:2010cf,Martinez:2010sc,Ryblewski:2010bs,Martinez:2010sd,Ryblewski:2011aq,Florkowski:2011jg,Martinez:2012tu,Ryblewski:2012rr,Florkowski:2012as,Bazow:2013ifa,Strickland:2013uga,Florkowski:2013uqa,Tinti:2013vba,Strickland:2014eua,Florkowski:2014txa,Florkowski:2014bba}%
. While second-order hydrodynamics is constructed from an expansion around a
local equilibrium state, anisotropic hydrodynamics originates from an
expansion around a dynamically-evolving anisotropic background.

So far, fluid-dynamical theories that include only the effects of shear
viscous corrections have been considered sufficient to describe the strongly
interacting system created in ultrarelativistic heavy ion collisions. 
However, since QCD is a nonconformal field theory one should not neglect the bulk 
viscous corrections to the ideal energy momentum tensor if one wants a complete and self-consistent description of the dynamics. 
While the accuracy of second-order and anisotropic
hydrodynamics has been investigated in the conformal and/or
massless limits \cite{Florkowski:2013lza,Florkowski:2013lya}, there have not
been comparisons of complete second-order formulations in the 
nonconformal case.  It was recently shown that
Israel-Stewart theory \cite{Israel:1979wp}, which is the most widespread formulation
of relativistic dissipative fluid dynamics, is not able to reproduce exact
solutions of the \textit{massive} 0+1d Boltzmann equation in the relaxation
time approximation \cite{Florkowski:2014sfa}.  Therefore, one is led to 
ask whether more complete formulations of second-order viscous hydrodynamics can better reproduce
the exact solution.

In the last months, some progress has been made in the second-order viscous
hydrodynamics framework within the 14-moment approximation \cite%
{Denicol:2014vaa} and in the anisotropic hydrodynamics framework \cite%
{Nopoush:2014pfa}. Both of these formalisms have been extended to provide a
more accurate description of massive and, consequently, nonconformal,
systems. In Ref.~\cite{Nopoush:2014pfa} it was shown that inclusion of an
explicit bulk degree of freedom in the anisotropic hydrodynamics framework
results in a quite reasonable agreement with the exact kinetic solutions. In
this paper we take another step in this direction and compare the solutions of second-order hydrodynamics obtained using the 14-moment
approximation \cite{Denicol:2014vaa} with the exact kinetic solution from 
Ref.~\cite{Florkowski:2014sfa}. We demonstrate that the failure of Israel-Stewart
theory in reproducing solutions of the Boltzmann equation in the massive
case occurs because this theory does not take into account the coupling
between bulk viscous pressure and the shear-stress tensor. We find that, for the case of the bulk viscous pressure, such coupling terms become as important as
the corresponding first-order Navier-Stokes term and must be included in
order to obtain a reasonable agreement with the microscopic theory. This
indicates that the coupling between the two viscous contributions can be
relevant in the description of nonconformal fluids.

We further compare the recent solutions of anisotropic hydrodynamics derived
in Ref.~\cite{Nopoush:2014pfa} with those of second order viscous
hydrodynamics. We find that both are able to reproduce the exact solution
with comparable accuracy. Such good agreement found between anisotropic
hydrodynamics and solutions of the Boltzmann equation is encouraging since
in anisotropic hydrodynamics the shear-bulk couplings do not need to be
included explicitly, but are instead implicit in the formalism.

The structure of this paper is as follows. In Sec.~\ref{sect:14moment} we
present the recently obtained second-order viscous hydrodynamics equations
of motion obtained using the 14-moment approximation. In Sec.~\ref%
{sect:ahydro} we present the necessary 0+1d anisotropic hydrodynamics
equations including the bulk degree of freedom. In Sec.~\ref{sect:exact} we
briefly review the method for solving the 0+1d massive Boltzmann equation
exactly. In Sec.~\ref{sect:results} we present our numerical results. In
Sec.~\ref{sect:conc} we present our conclusions and an outlook for the
future. In the appendix, we collect expressions for the necessary
thermodynamic integrals and their asymptotic expansions.

\subsection*{Notation and conventions}

We use natural units with $\hbar = c = k_B = 1$. The metric tensor has the
form $g^{\mu\nu} = \mathrm{diag}(1,-1,-1,-1)$. The spacetime coordinates are
denoted as $x^\mu=(t,x,y,z)$, and the longitudinal proper time is $\tau=%
\sqrt{t^2-z^2}$.

\subsection*{Bulk variables}
\label{subsect:bulkvars}

In order to compare the various approximation schemes considered herein, we will specialize in the end to the case that the equilibrium distribution is a classical massive Boltzmann distribution.  In this case, the isotropic equilibrium bulk variables are
\begin{eqnarray}
n_{\rm eq}(T,m) &=& 4 \pi \tilde{N} T^3 \, \hat{m}_{\rm eq}^2 K_2\left( \hat{m}_{\rm eq}\right) \,  , \label{eq:neq} \\
{\cal S}_{\rm eq}(T,m) &=& 4 \pi \tilde{N} T^3 \, \hat{m}_{\rm eq}^2 
 \Big[ 4 K_{2}\left( \hat{m}_{\rm eq} \right) + \hat{m}_{\rm eq} K_{1} \left( \hat{m}_{\rm eq}\right) \Big] \, ,
\label{eq:sigmaeq} \\
{\cal E}_{\rm eq}(T,m) &=& 4 \pi \tilde{N} T^4 \, \hat{m}_{\rm eq}^2
 \Big[ 3 K_{2}\left( \hat{m}_{\rm eq} \right) + \hat{m}_{\rm eq} K_{1} \left( \hat{m}_{\rm eq} \right) \Big] \, , 
\label{eq:epsiloneq} \\
{\cal P}_{\rm eq}(T,m) &=& n_{\rm eq}(T,m) \, T \, ,
\label{eq:Peq}
\end{eqnarray}
\checked{rs}
where $n_{\rm eq}$, ${\cal S}_{\rm eq}$, ${\cal E}_{\rm eq}$, ${\cal P}_{\rm eq}$ are the equilibrium number density, entropy density, energy density, pressure, respectively, $\hat{m}_{\rm eq} = m/T$, and $\tilde{N} = N_{\rm dof}/(2\pi)^3$ with $N_{\rm dof}$ being the number of degrees of freedom.

\section{14-moment approximation approach to fluid dynamics}
\label{sect:14moment} 

Using the 14-moment approximation one can derive the equations of motion for
a relativistic fluid from the relativistic Boltzmann kinetic equation. In
this way, the continuity equations for the energy-momentum tensor 
\begin{equation}
\partial _{\mu }T^{\mu \nu }=0\,,  \label{enmomcon}
\end{equation}
have to be solved together with the relaxation-type equations for the bulk
viscous pressure $\Pi $ and the shear-stress tensor $\pi ^{\mu \nu }$ \cite%
{Denicol:2014vaa}, 
\begin{eqnarray}
\tau _{\Pi }\dot{\Pi}+\Pi &=&-\zeta \theta -\delta _{\Pi \Pi }\Pi \theta
+\varphi _{\mathrm{1}}\Pi ^{2}+\lambda _{\Pi \pi }\pi ^{\mu \nu }\sigma
_{\mu \nu }+\varphi _{\mathrm{3}}\pi ^{\mu \nu }\pi _{\mu \nu } \, ,
\label{dnmreq1a} \\
\tau _{\pi }\dot{\pi}^{\langle \mu \nu \rangle }+\pi ^{\mu \nu } &=&2\eta
\sigma ^{\mu \nu }+2\tau _{\pi }\pi _{\alpha }^{\langle \mu }\omega ^{\nu
\rangle \alpha }-\delta _{\pi \pi }\pi ^{\mu \nu }\theta +\varphi _{\mathrm{7%
}}\pi _{\alpha }^{\langle \mu }\pi ^{\nu \rangle \alpha }-\tau _{\pi \pi
}\pi _{\alpha }^{\langle \mu }\sigma ^{\nu \rangle \alpha }  \notag \\
&&+\,\lambda _{\pi \Pi }\Pi \sigma ^{\mu \nu }+\varphi _{6}\Pi \pi ^{\mu \nu
}\,.  \label{dnmreq1b}
\end{eqnarray}
Here we have neglected the effect of net-charge diffusion. Above, we
introduced the vorticity tensor $\omega _{\mu \nu }\equiv (\nabla _{\mu
}u_{\nu }-\nabla _{\nu }u_{\mu })/2$, the shear tensor \mbox{$\sigma_{\mu%
\nu}\equiv \nabla_{\langle\mu} u_{\nu\rangle}$} and the expansion scalar $%
\theta \equiv \nabla _{\mu }u^{\mu }$, where $u^{\mu }$ is the fluid
four-velocity and $\nabla _{\mu }\equiv \Delta _{\mu }^{\nu }\partial _{\nu
} $ is the projected spatial gradient. We use the notation $A^{\langle \mu
\nu \rangle }\equiv \Delta _{\alpha \beta }^{\mu \nu }A^{\alpha \beta }$,
with $\Delta _{\alpha \beta }^{\mu \nu }\equiv (\Delta _{\alpha }^{\mu
}\Delta _{\beta }^{\nu }+\Delta _{\beta }^{\mu }\Delta _{\alpha }^{\nu
}-2/3\Delta ^{\mu \nu }\Delta _{\alpha \beta })/2$, where $\Delta ^{\mu \nu
}\equiv g^{\mu \nu }-u^{\mu }u^{\nu }$. In Eqs.~(\ref{dnmreq1a}) and (\ref%
{dnmreq1b}) we have also introduced the shorthand notation for the
proper-time derivative $\dot{(\,\,)}\equiv d/d\tau $.

The terms multiplying different tensor structures in (\ref{dnmreq1a}) and (%
\ref{dnmreq1b}) are transport coefficients. They are complicated functions
of temperature and the particle's mass, and their form should be found by
matching (\ref{dnmreq1a}) and (\ref{dnmreq1b}) with the underlying
microscopic theory. As shown in Ref. \cite{Molnar:2013lta}, the terms $%
\varphi _{\mathrm{1}}\Pi ^{2}$, $\varphi _{\mathrm{3}}\pi ^{\mu \nu }\pi
_{\mu \nu }$, $\varphi _{6}\Pi \pi ^{\mu \nu }$, and $\varphi _{\mathrm{7}%
}\pi _{\alpha }^{\langle \mu }\pi ^{\nu \rangle \alpha }$ appear only
because the collision term is nonlinear in the single-particle distribution
function. In the case of the relaxation time approximation, that will be
employed throughout this paper, the collision term is assumed to be linear
in the nonequilibrium single-particle distribution function and one can
explicitly show that $\varphi _{\mathrm{1}}=\varphi _{\mathrm{3}}=\varphi _{%
\mathrm{6}}=\varphi _{\mathrm{7}}=0${.} One should stress, however, that
Eqs.~(\ref{dnmreq1a}) and (\ref{dnmreq1b}) include a coupling between the
shear and bulk relaxation equations (the terms $\lambda _{\Pi \pi }$ and $%
\lambda _{\pi \Pi }$), which are absent in the traditional Israel-Stewart
viscous hydrodynamics.  One can find a plot
of the various transport coefficients in Fig.~1 of Ref.~\cite%
{Denicol:2014vaa}.  There have been some prior works
that have considered shear-bulk couplings in viscous hydrodynamics,
see e.g. Refs. \cite{Tsumura:2009vm,Monnai:2010qp,Monnai:2012jc}, but
for the most part, the existence of these types of couplings has been
ignored in the literature.

In the 0+1d case describing one-dimensional and boost invariant
expansion, the formulae~(\ref{enmomcon})--(\ref{dnmreq1b}) reduce to
\begin{eqnarray}
\dot{\mathcal{E}} &=&-\frac{\mathcal{E}+\mathcal{P}+\Pi -\pi }{\tau }\,,
\label{dnmreq2a} \\
\tau _{\Pi }\dot{\Pi}+\Pi &=&-\frac{\zeta }{\tau }-\delta _{\Pi \Pi }\frac{%
\Pi }{\tau }+\lambda _{\Pi \pi }\frac{\pi }{\tau }\,,  \label{dnmreq2b} \\
\tau _{\pi }\dot{\pi}+\pi &=&\frac{4}{3}\frac{\eta}{\tau} -\left( \frac{1}{3}\tau
_{\pi \pi }+\delta _{\pi \pi }\right) \frac{\pi }{\tau }+\frac{2}{3}\lambda _{\pi \Pi }%
 \frac{\Pi}{\tau}  \,,  \label{dnmreq2c}
\end{eqnarray}
where $\mathcal{E}$ and $\mathcal{P}$ are the energy density and
thermodynamic pressure, respectively. We note that in 0+1d the vorticity
tensor vanishes and the term $2\tau _{\pi }\pi_{\alpha }^{\langle \mu
}\omega ^{\nu \rangle \alpha }$ has no effect on the dynamics of the fluid.
The coefficients appearing in the equation for the bulk pressure are the
following 
\begin{eqnarray}
\frac{\zeta }{\tau _{\Pi }} &=&\left( \frac{1}{3}-c_{s}^{2}\right) (\mathcal{%
E}+\mathcal{P})-\frac{2}{9}(\mathcal{E}-3\mathcal{P})-\frac{m^{4}}{9}%
I_{-2,0}\,, \\
\frac{\delta _{\Pi \Pi }}{\tau _{\Pi }} &=&1-c_{s}^{2}-\frac{m^{4}}{9}\gamma
_{\mathrm{2}}^{\mathrm{(0)}}\,, \\
\frac{\lambda _{\Pi \pi }}{\tau _{\Pi }} &=&\frac{1}{3}-c_{s}^{2}+\frac{m^{2}%
}{3}\gamma _{\mathrm{2}}^{\mathrm{(2)}}\,.
\end{eqnarray}
On the other hand, the coefficients in the equation for the shear pressure
are 
\begin{eqnarray}
\frac{\eta }{\tau _{\pi }} &=&\frac{4}{5}\mathcal{P}+\frac{1}{15}(\mathcal{E}%
-3\mathcal{P})-\frac{m^{4}}{15}I_{-2,0}\,, \\
\frac{\delta _{\pi \pi }}{\tau _{\pi }} &=&\frac{4}{3}+\frac{1}{3}%
m^{2}\gamma _{\mathrm{2}}^{\mathrm{(2)}}\,, \\
\frac{\tau _{\pi \pi }}{\tau _{\pi }} &=&\frac{10}{7}+\frac{4}{7}m^{2}\gamma
_{\mathrm{2}}^{\mathrm{(2)}}\,, \\
\frac{\lambda _{\pi \Pi }}{\tau _{\pi }} &=&\frac{6}{5}-\frac{2}{15}%
m^{4}\gamma _{\mathrm{2}}^{\mathrm{(0)}}\,,
\end{eqnarray}
where we have introduced the sound velocity squared
\begin{equation}
c_{s}^{2}=\frac{\mathcal{E}+\mathcal{P}}{\beta _{0}I_{\mathrm{3,0}}}\,,
\end{equation}
and $\beta _{0}=I_{\mathrm{1,0}}/\mathcal{P}$. The coefficients $\gamma
_{n}^{\mathrm{(0)}}$ and $\gamma _{n}^{\mathrm{(2)}}$ are complicated
functions of $T$ and $m$ given by 
\begin{eqnarray}
\gamma _{n}^{\mathrm{(0)}}
&=&(E_{0}+B_{0}m^{2})I_{-n,0}+D_{0}I_{1-n,0}-4B_{0}I_{2-n,0}\,, \\
\gamma _{n}^{\mathrm{(2)}} &=&\frac{I_{4-n,2}}{I_{4,2}}\,,
\end{eqnarray}
where 
\begin{eqnarray}
\frac{D_{0}}{3B_{0}} &=&-4\frac{I_{3,1}I_{2,0}-I_{4,1}I_{1,0}}{%
I_{3,0}I_{1,0}-I_{2,0}I_{2,0}}\equiv -C_{2}\,, \\
\frac{E_{0}}{3B_{0}} &=&m^{2}+4\frac{I_{3,1}I_{3,0}-I_{4,1}I_{2,0}}{%
I_{3,0}I_{1,0}-I_{2,0}I_{2,0}}\equiv -C_{1}\,, \\
B_{0} &=&-\frac{1}{3C_{1}I_{2,1}+3C_{2}I_{3,1}+3I_{4,1}+5I_{4,2}}\,.
\end{eqnarray}
Here we make use of the thermodynamic functions $I_{n,q}$ defined by the
integrals 
\begin{equation}
I_{n,q}(T,m)=\frac{1}{(2q+1)!!}\int dK(u_{\mu }k^{\mu })^{n-2q}(-\Delta
_{\mu \nu }k^{\mu }k^{\nu })^{q}f_{0k}\,,
\label{intI}
\end{equation}
where, herein, the equilibrium distribution function is assumed to be a classical Boltzmann
distribution \mbox{$f_{0k} = \exp\left(-u_\mu k^\mu/T\right)$}, and the integration
measure is $dK=N_{\mathrm{dof}}\,d^{3}k/((2\pi )^{3}k^{0})$. The relevant
integrals $I_{n,q}(T,m)$ are expressed in terms of special functions in
Appendix \ref{sect:integrals}. Finally, in this paper we only consider the
Boltzmann equation in the relaxation time approximation. In this case,
the shear and bulk relaxation times, $\tau _{\pi }$ and $\tau _{\Pi }$,
respectively, are equal to the microscopic relaxation time $\tau _{\mathrm{eq%
}}$, i.e., $\tau _{\Pi }=\tau _{\pi }=\tau _{\mathrm{eq}}$   
\cite{Denicol:2014vaa}~\footnote{In a more general case
the relaxation times for the shear and bulk pressures may differ from each other.}.

We note here that there are other formulations of second-order hydrodynamics
which have different values for the various transport coefficients listed
above. For example, if one uses the naive Israel-Stewart theory one has $%
\tau _{\pi \pi }=0$. In addition, even using the method of moments one finds
that the coefficients depend on the number of moments considered. For
example, for a gas of massless particles with constant cross sections, one
has $\tau _{\pi \pi }=134/77$ in the 23-moment approximation and $\tau _{\pi
\pi }\simeq 1.69$ in the 32- and 41-moment approximations \cite%
{Denicol:2012cn}. In this paper, we will use the coefficients calculated in
the 14-moment approximation using the relaxation time approximation \cite%
{Denicol:2014vaa}. We note that, for the case of the relaxation time
approximation, the transport coefficients of hydrodynamics have not been
calculated beyond the 14-moment approximation.

Finally, in this context we note that even the form of the bulk pressure
evolution equations put forward by different authors are different and
generally speaking until the recent papers of Denicol et al. no authors
included explicit shear-bulk couplings, see e.g. \cite%
{Israel:1976tn,Israel:1979wp,Muronga:2003ta,Heinz:2005bw,Romatschke:2009im,Song:2009rh,Jaiswal:2013npa}%
. As shown in Ref.~\cite{Florkowski:2014sfa}, approaches that do not
explicitly include the shear-bulk couplings do not agree with the bulk
pressure evolution obtained via exact solution of the Boltzmann equation.
The complete 14-moment second-order viscous hydrodynamics equations presented
in this section include shear-bulk couplings, $\lambda _{\pi \Pi }$ and $%
\lambda _{\Pi \pi }$. As we will see in the results section, including these
couplings results in much better agreement with the exact kinetic solution
compared to Israel-Stewart second-order viscous hydrodynamics.

\section{Anisotropic hydrodynamics approach}
\label{sect:ahydro} 

Anisotropic hydrodynamics (aHydro) is an alternative framework for obtaining
the necessary non-equilibrium evolution equations. In contrast to
traditional viscous hydrodynamics approaches, which make an expansion around
the equilibrium state, anisotropic hydrodynamics expands the underlying
distribution function around a momentum-space anisotropic state. In this way, the
potentially large degree of momentum-space anisotropy in the system is
included in the leading order of expansion and treated non-perturbatively.
Moreover, in this framework one is not restricted by the condition of being
close to the equilibrium state since the dynamical background is allowed to
possess even large momentum-space anisotropies.

In its newest formulation \cite{Nopoush:2014pfa}, the framework of
anisotropic hydrodynamics allows for a degree of freedom associated with the
bulk pressure of the system. This is accomplished using the following form
for the underlying distribution function 
\begin{equation}
f(x,p) = f_{\mathrm{iso}}\!\left(\frac{1}{\lambda}\sqrt{p_\mu \Xi^{\mu\nu}
p_\nu}\right) ,  \label{eq:genf}
\end{equation}
with $\Xi^{\mu\nu} = u^\mu u^\nu + \xi^{\mu\nu} - \Delta^{\mu\nu} \Phi$,
where $u^\mu$ is the four-velocity associated with the local rest frame, $%
\xi^{\mu\nu}$ is a symmetric and traceless tensor, and $\Phi$ is the bulk
degree of freedom. The quantities $u^\mu$, $\xi^{\mu\nu}$, and $\Phi$ are
understood to be functions of space and time and obey $u^\mu u_\mu = 1$, ${%
\xi^{\mu}}_\mu = 0$, and $u_\mu \xi^{\mu\nu} = 0$.

Taking the isotropic distribution $f_{\mathrm{iso}}$ to be a Boltzmann
distribution and assuming 0+1d boost-invariant evolution, the dynamics of
the system is determined by the three aHydro equations \cite{Nopoush:2014pfa}
\begin{eqnarray}
&&\partial _{\tau }\log \alpha _{x}^{2}\alpha _{z}+\left[ 3+\hat{m}\frac{%
K_{1}(\hat{m})}{K_{2}(\hat{m})}\right] \,\partial _{\tau }\log \lambda +%
\frac{1}{\tau }=\frac{1}{\tau _{\mathrm{eq}}}\left[ \frac{1}{\alpha
_{x}^{2}\alpha _{z}}\frac{T}{\lambda }\frac{K_{2}(\hat{m}_{\mathrm{eq}})}{%
K_{2}(\hat{m})}-1\right] \,,  \label{eq:final0m} \\
&&\left( 4\tilde{\mathcal{H}}_{3}-\tilde{\Omega}_{m}\right) \partial _{\tau
}\log \lambda +\tilde{\Omega}_{T}\partial _{\tau }\log \alpha _{x}^{2}+%
\tilde{\Omega}_{L}\partial _{\tau }\log \alpha _{z}=-\frac{1}{\tau }\tilde{%
\Omega}_{L}\,,  \label{eq:final1m} \\
&&\partial _{\tau }\log \left( \frac{\alpha _{x}}{\alpha _{z}}\right) -\frac{%
1}{\tau }+\frac{3}{4\tau _{\mathrm{eq}}}\frac{\xi _{z}}{\alpha
_{x}^{2}\alpha _{z}}\left( \frac{T}{\lambda }\right) ^{2}\frac{K_{3}(\hat{m}%
_{\mathrm{eq}})}{K_{3}(\hat{m})}=0\,,  \label{eq:final2m}
\end{eqnarray}%
where $\tau _{\mathrm{eq}}$ is the microscopic relaxation time. Above the
variables $\alpha _{i}$ are a particular combination of the traceless ($\xi
_{i}$) and traceful component ($\Phi $) parts of the underlying
momentum-space anisotropy tensor with $\alpha _{i}=(1+\xi _{i}+\Phi )^{-1/2}$%
. The variable $\lambda $ is the non-equilibrium energy scale in the
distribution function (\ref{eq:genf}) and we have defined two dimensionless
mass scales $\hat{m}_{\mathrm{eq}}=m/T$ and $\hat{m}=m/\lambda $. The
integrals $\tilde{\mathcal{H}}_{3}$, $\tilde{\Omega}_{T}$, $\tilde{\Omega}%
_{L}$, and $\tilde{\Omega}_{m}$ are defined in Eqs.~(A3) and (59) of Ref.~%
\cite{Nopoush:2014pfa}.

Equations (\ref{eq:final0m})--(\ref{eq:final2m}) determine the proper-time
evolution of $\alpha _{x}$, $\alpha _{z}$, and $\lambda $. The temperature,
or more accurately, the effective temperature appearing above is determined
by requiring energy conservation at all proper times. This results in the
dynamical Landau matching condition 
\begin{equation}
\tilde{\mathcal{H}}_{3}\lambda ^{4}=4\pi \tilde{N}T^{4}\hat{m}_{\mathrm{eq}%
}^{2}\Big[3K_{2}\left( \hat{m}_{\mathrm{eq}}\right) +\hat{m}_{\mathrm{eq}%
}K_{1}\left( \hat{m}_{\mathrm{eq}}\right) \Big].  \label{eq:dlm}
\end{equation}%
When the system is transversely homogeneous, the longitudinal and transverse
pressure can be expressed as 
\begin{eqnarray}
\mathcal{P}_{T} &=&\tilde{\mathcal{H}}_{3T}({\boldsymbol{\xi }},\Phi ,\hat{m}%
)\,\lambda ^{4}\,, \\
\mathcal{P}_{L} &=&\tilde{\mathcal{H}}_{3L}({\boldsymbol{\xi }},\Phi ,\hat{m}%
)\,\lambda ^{4}\,,  \label{plptah}
\end{eqnarray}
where ${\boldsymbol{\xi }} = (\xi_x,\xi_y,\xi_z)$ are the diagonal components
of the $\xi^{\mu\nu}$.  The $\tilde{\mathcal{H}}_{3T}$ and $\tilde{\mathcal{H}}_{3L}$
functions appearing above are defined by Eqs.~(A8) and (A13) in Ref.~\cite{Nopoush:2014pfa}.
Likewise, the bulk pressure can be computed using 
\begin{equation}
\Pi (\tau )=\frac{1}{3}\left[ \mathcal{P}_{L}(\tau )+2\mathcal{P}_{T}(\tau
)-3\mathcal{P}(\tau )\right] .  \label{eq:PIkz}
\end{equation}


\section{Exact solutions of Boltzmann kinetic equation in the relaxation
time approximation}

\label{sect:exact} 

Herein we focus on a transversely homogeneous boost-invariant system. In
this case the hydrodynamic flow $u^\mu$ should have the Bjorken form in the
lab frame $u^\mu = \left(t/\tau,0,0,z/\tau\right)$~\cite{Bjorken:1982qr}.
This implies that the distribution function $f(x,p)$ can depend only on $%
\tau $, $w$, and $p_T$ with~\cite{Bialas:1984wv,Bialas:1987en} $w = t p_L -
z E$. Using $w$ and $p_L$ one can define another boost-invariant variable $v
= Et-p_L z = \sqrt{w^2+\left( m^2+\vec{p}_T^{\,\,2}\right) \tau^2}$. Using
the boost-invariant variables introduced above, the relaxation time approximation
kinetic equation may be
written in a simple form 
\begin{equation}
\frac{\partial f}{\partial \tau} = \frac{f_{\mathrm{eq}}-f}{\tau_{\mathrm{eq}%
}} \, ,  \label{eq:feq}
\end{equation}
where the boost-invariant form of the equilibrium distribution function is 
\begin{equation}
f_{\mathrm{eq}}(\tau, w, p_T) = \exp\left[ - \frac{\sqrt{w^2+ \left(
m^2+p_T^2 \right) \tau^2}}{T(\tau) \tau} \, \right] .  \label{eq:geq}
\end{equation}
The general form of solutions of Eq.~(\ref{eq:feq}) can be expressed as~\cite%
{Florkowski:2013lza,Florkowski:2013lya,Florkowski:2014sfa,Baym:1984np,Baym:1985tna,Heiselberg:1995sh,Wong:1996va}
\begin{equation}
f(\tau,w,p_T) = D(\tau,\tau_0) f_0(w,p_T) + \int_{\tau_0}^\tau \frac{%
d\tau^\prime}{\tau_{\mathrm{eq}}(\tau^\prime)} \, D(\tau,\tau^\prime) \, f_{%
\mathrm{eq}}(\tau^\prime,w,p_T) \, ,  \label{eq:solG}
\end{equation}
where we have introduced the damping function 
\begin{equation}
D(\tau_2,\tau_1) = \exp\left[-\int_{\tau_1}^{\tau_2} \frac{%
d\tau^{\prime\prime}}{\tau_{\mathrm{eq}}(\tau^{\prime\prime})} \right] .
\label{eq:damping}
\end{equation}
For the purposes of this paper, we will assume that at $\tau=\tau_0$ the
distribution function $f$ can be expressed in spheroidal
Romatschke-Strickland form \cite{Romatschke:2003ms} with the underlying
Boltzmann distribution being an isotropic distribution 
\begin{equation}
f_0(w,p_T) = \exp\left[ -\frac{\sqrt{(1+\xi_0) w^2 + (m^2+p_T^2) \tau_0^2}}{%
\Lambda_0 \tau_0}\, \right] ,  \label{eq:G0}
\end{equation}
where $\xi_0$ measures the initial momentum-space anisotropy and $\Lambda_0$
is the initial spheroidal momentum-scale. This form simplifies to an
isotropic Boltzmann distribution if the anisotropy parameter $\xi_0$ is
zero, in which case the transverse momentum scale $\Lambda_0$ can be
identified with the system's initial temperature $T_0$. Using Eq.~(\ref%
{eq:solG}) one can derive an integral equation satisfied by the energy
density~\cite{Florkowski:2014sfa} 
\begin{eqnarray}
&& \hspace{-5mm} 2 m^2 T(\tau) \Bigl[ 3 T(\tau) K_2(\hat{m}_{\mathrm{eq}%
}(\tau)) + m K_1(\hat{m}_{\mathrm{eq}}(\tau)) \Bigr]  \notag \\
&& = D(\tau,\tau_0) \Lambda^4_0 \tilde{\mathcal{H}}_2\left[ \frac{\tau_0}{%
\tau \sqrt{1+\xi_0}},\frac{m}{\Lambda_0}\right] + \int\limits_{\tau_0}^\tau 
\frac{d\tau^\prime}{\tau_{\mathrm{eq}}(\tau^{\prime })} D(\tau,\tau^\prime)
T^4(\tau^\prime) \tilde{\mathcal{H}}_2\left[ \frac{\tau^\prime}{\tau},\hat{m}%
_{\mathrm{eq}}(\tau^{\prime })\right] ,  \label{eq:LM2}
\end{eqnarray}
where $T$ is the effective temperature which is related to the energy
density via Eq.~(\ref{eq:epsiloneq}). The function $\tilde{\mathcal{H}}%
_2(y,z)$ above is defined by the integral 
\begin{equation}
\tilde{\mathcal{H}}_2(y,z) = \int\limits_0^\infty du\, u^3 \, \mathcal{H}%
_2\left(y,\frac{z}{u} \right) \, \exp\left(-\sqrt{u^2+z^2}\right) ,
\label{eq:tildeH2}
\end{equation}
with 
\begin{equation}
\mathcal{H}_2(y,\zeta) = y \left( \sqrt{y^2+\zeta^2} + \frac{1+\zeta^2}{%
\sqrt{y^2-1}} \tanh^{-1} \sqrt{\frac{y^2-1}{y^2+\zeta^2}} \, \right) .
\label{eq:H2an}
\end{equation}

\begin{figure}[t]
\centerline{\includegraphics[angle=0,width=0.65%
\textwidth]{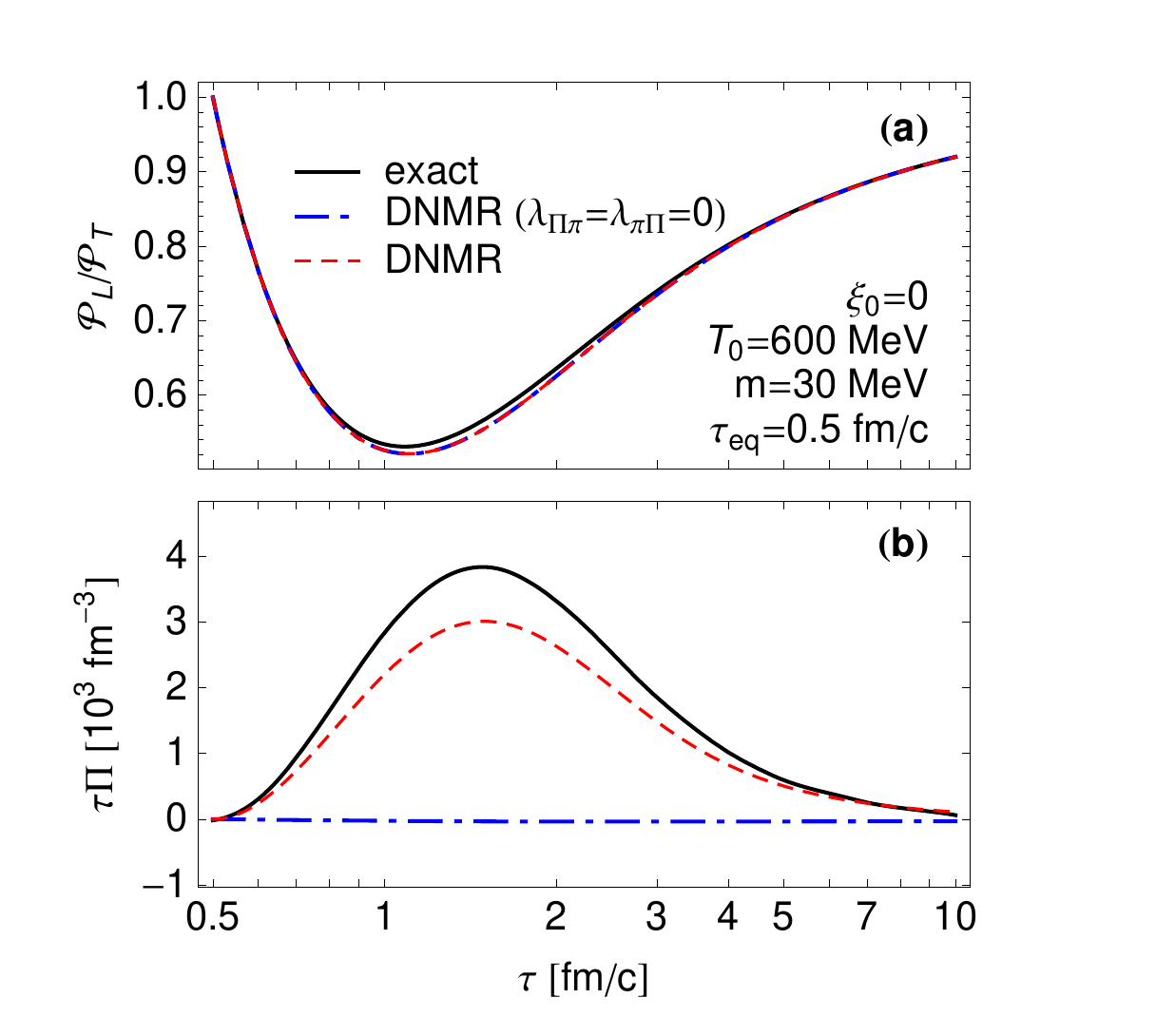}}
\caption{(Color online) Time evolution of the pressure anisotropy $\mathcal{P%
}_L/\mathcal{P}_T$ (a) and the bulk pressure (b). Three lines
describe three different results: the exact solution of the Boltzmann
equation \protect\cite{Florkowski:2014sfa} (black solid line), the result of
the full second-order viscous hydrodynamics \protect\cite{Denicol:2014vaa}
including the shear-bulk couplings $\protect\lambda_{\Pi\protect\pi}$ and $%
\protect\lambda_{\protect\pi\Pi}$ (red dashed line), and the result of the
second-order hydrodynamics with $\protect\lambda_{\Pi\protect\pi}=\protect%
\lambda_{\protect\pi\Pi}=0$ (blue dot-dashed line). For both panels we use $%
m=$ 30 MeV, $\protect\tau_0$ = 0.5 fm/c, $\protect\tau_{\mathrm{eq}} =%
\protect\tau_{\protect\pi} =\protect\tau_{\Pi}$ = 0.5 fm/c, and $T_0$ = 600
MeV. The initial spheroidal anisotropy parameter fixing the initial
distribution function equals $\protect\xi_0=0$, correspondingly, we use $%
\protect\pi_0=0$ and $\Pi_0$ = 0.}
\label{fig:coupl1}
\end{figure}

\begin{figure}[t]
\centerline{
\hspace{1cm}
\includegraphics[angle=0,width=0.575\textwidth]{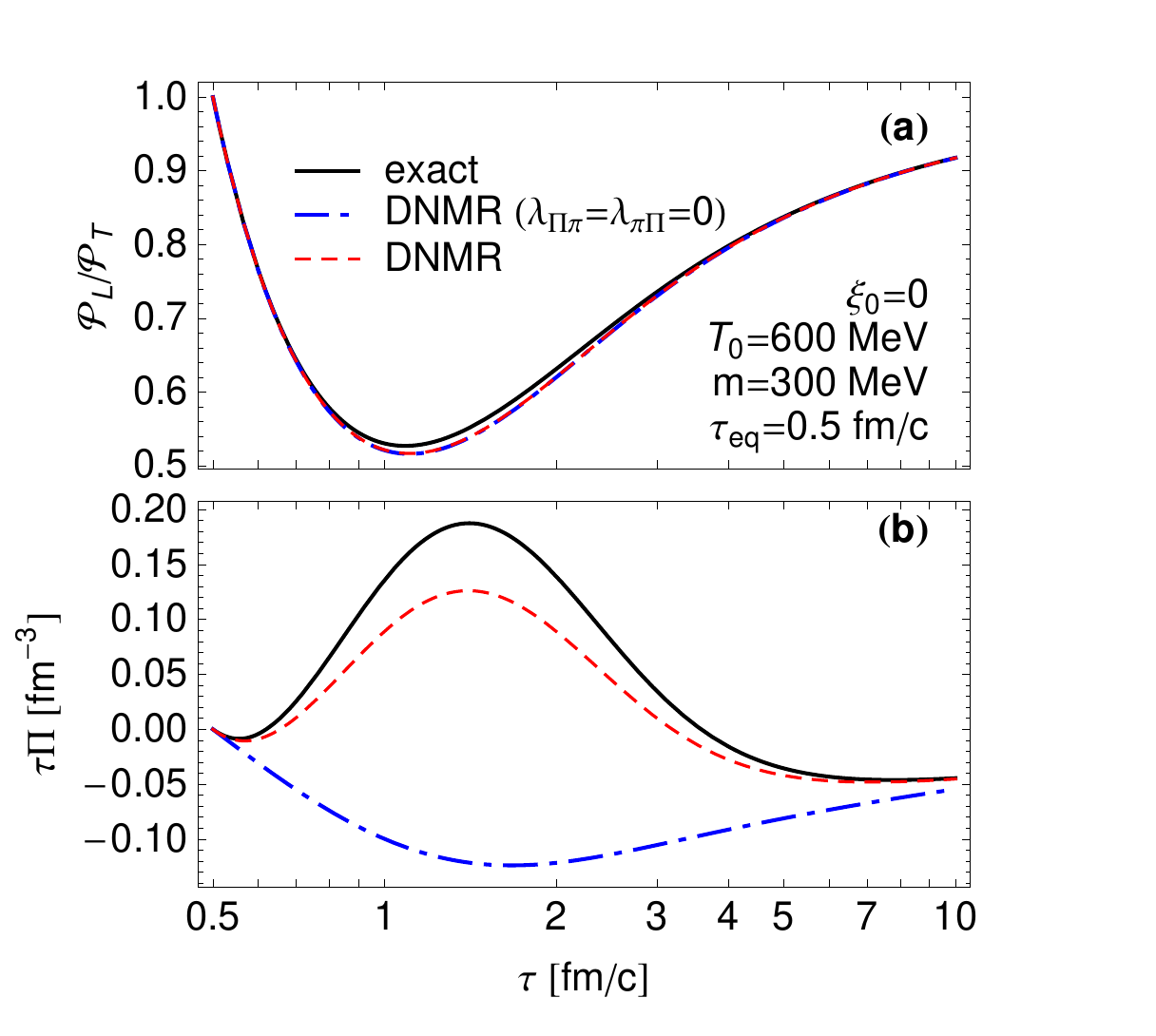}
\hspace{-1.5cm}
\includegraphics[angle=0,width=0.575\textwidth]{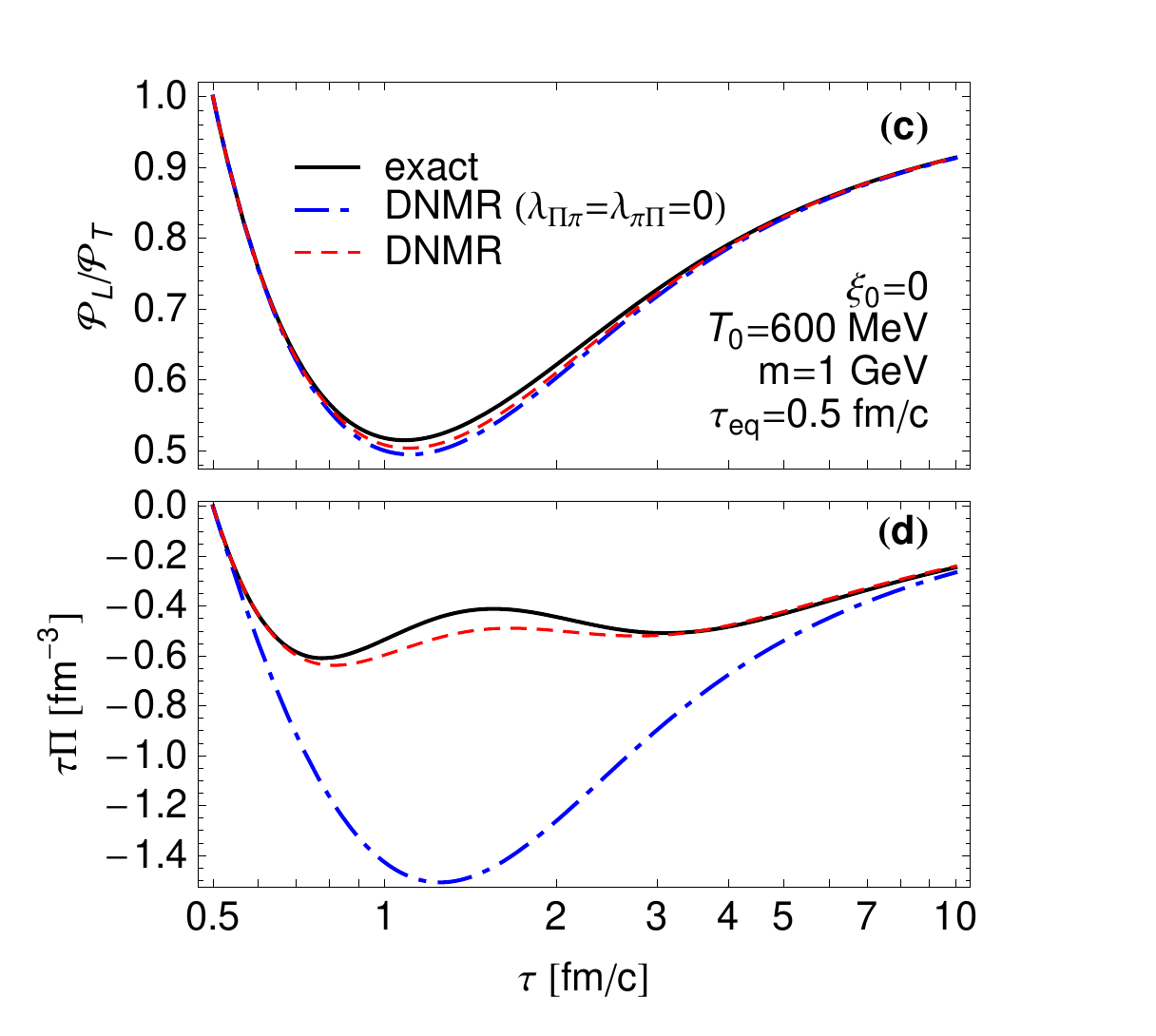}
}
\caption{(Color online) Same as Fig.~\protect\ref{fig:coupl1}
except here we take $m=300$ MeV ((a) and (b)) and $m=1$ GeV ((c) and (d)).}
\label{fig:coupl2}
\end{figure}

Equation~(\ref{eq:LM2}) can be solved numerically using the method of
iteration. Using this method, one makes an initial guess for the proper time
dependence of the effective temperature, e.g. the ideal hydrodynamics, plugs
this into the right hand side of Eq.~(\ref{eq:LM2}), and then one solves for
the effective temperature necessary to make the left and right hand sides
equal using a root finder. The resulting effective temperature
\textquotedblleft profile\textquotedblright\ is then used as the new
\textquotedblleft initial guess\textquotedblright\ and one repeats this
process iteratively until the effective temperature profile converges to a
given accuracy within the proper-time interval of interest. Once the
effective temperature is determined via iterative solution, one can use this
to determine the transverse pressure, longitudinal pressure, full
distribution function, etc. For further details, we refer the reader to Ref.~%
\cite{Florkowski:2014sfa}.


\section{Results}

\label{sect:results} 

In this Section we present and discuss our results for the proper-time
evolution of the system using different approaches. The results obtained
within 14-moment second-order viscous hydrodynamics and anisotropic hydrodynamics are
compared with the exact solution. We begin by emphasizing the importance of
including the full set of kinetic coefficients in the second-order viscous
relativistic hydrodynamics in order to describe the bulk pressure evolution
obtained via exact solution of the RTA (relaxation time approximation) Boltzmann equation. In order to
illustrate this point, below we will compare the second-order viscous
hydrodynamics predictions with and without the shear-bulk couplings. After
making this point, we then compare the results obtained using full
second-order viscous hydrodynamics with those obtained within the
anisotropic hydrodynamics. We find that the two approaches reproduce the
exact solution with comparable accuracy. In all cases we use a fixed
relaxation time of $\tau_{\mathrm{eq}} = \tau_\pi = \tau_\Pi = 0.5$ fm/c, an
initial time of $\tau_0 = 0.5$ fm/c, and an initial temperature of $T_0$ =
600 MeV. We will consider two different initial pressure anisotropies
corresponding to an isotropic initial condition ($\xi_0 = 0$) and a highly
oblate initial anisotropy ($\xi_0 = 100$). For the particle mass, we will
consider three different cases corresponding to $m = $ 30 MeV, 300 MeV, and
1 GeV.


\subsection{Shear-bulk couplings in the second-order viscous hydrodynamics}

\label{subsect:couplings} 

In this Section we solve the second-order hydrodynamic equations discussed
in Sec.~\ref{sect:14moment} and compare the obtained solutions with the
exact results. In order to have an overlap with our previous results
available in the literature, the initial temperature at $\tau_0 = 0.5$ fm/c
has been set equal to $T_0=$ 600 MeV. By solving the kinetic equation in the
relaxation time approximation using Eq.~(\ref{eq:LM2}) we obtain the
effective temperature $T(\tau)$. As mentioned previously, knowing $T(\tau)$
we can then calculate the exact pressures $\mathcal{P}_L(\tau)$ and $%
\mathcal{P}_T(\tau)$. We then use Eq.~(\ref{eq:PIkz}) to obtain the exact
bulk pressure $\Pi(\tau)$. The exact bulk pressure computed in this manner
can be compared directly with the second-order hydrodynamic result for $%
\Pi(\tau)$, which follows from Eqs.~(\ref{dnmreq2a})--(\ref{dnmreq2c}).

In addition, the second-order hydrodynamics results for $\mathcal{P}$, $\Pi$%
, and $\pi$ can be used to determine $\mathcal{P}_T $ and $\mathcal{P}_L$
via 
\begin{eqnarray}
\mathcal{P}_L &=& \mathcal{P} + \Pi - \pi \, ,  \notag \\
\mathcal{P}_T &=& \mathcal{P} + \Pi + \pi/2 \, ,  \label{eq:PTPL}
\end{eqnarray}

\begin{figure}[t]
\centerline{\includegraphics[angle=0,width=0.65%
\textwidth]{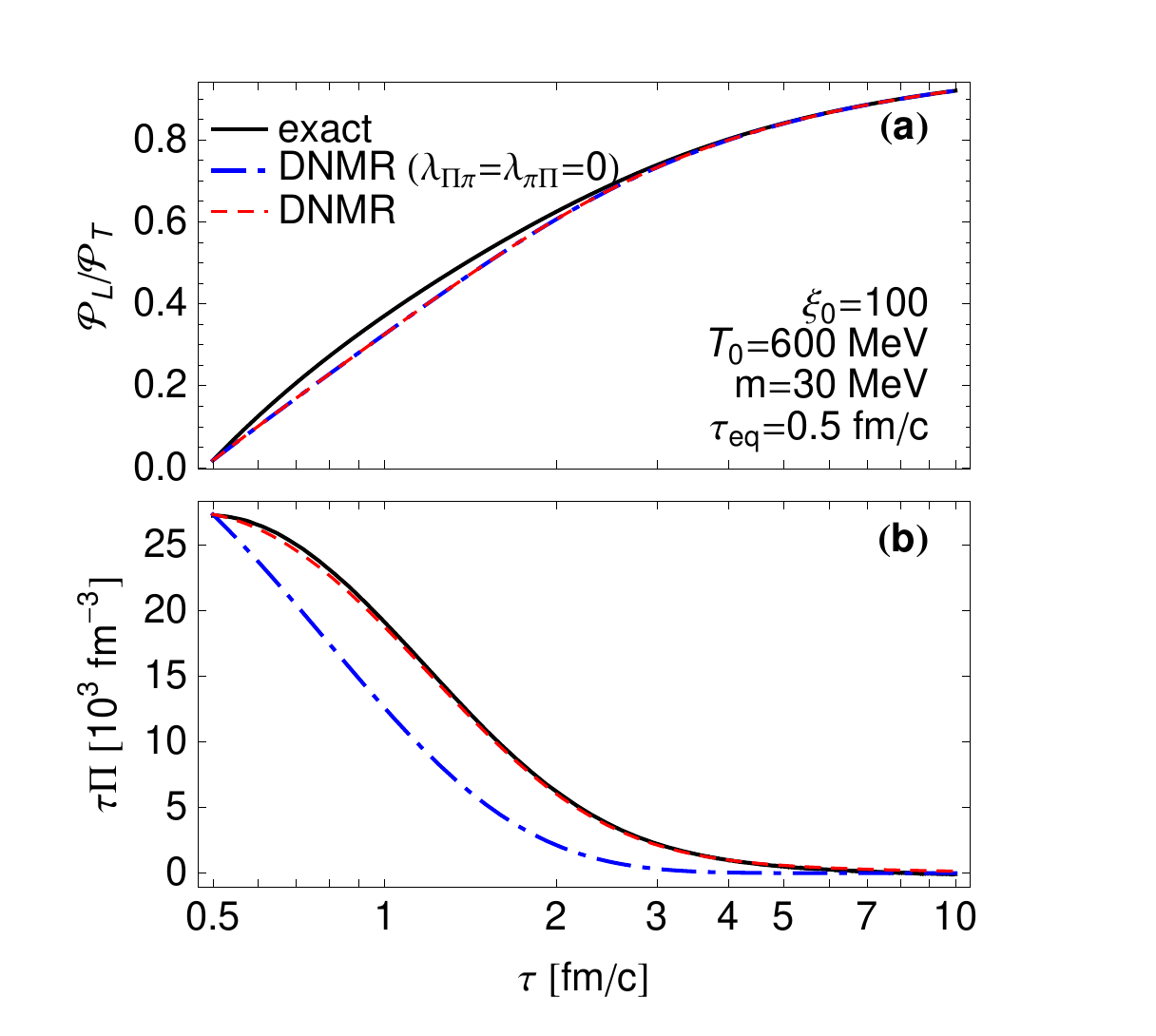}}
\caption{(Color online) Same as Fig.~\protect\ref{fig:coupl1} except here we
take $\protect\xi_0=100$.}
\label{fig:coupl3}
\end{figure}

\begin{figure}[t]
\centerline{
\hspace{1cm}
\includegraphics[angle=0,width=0.575\textwidth]{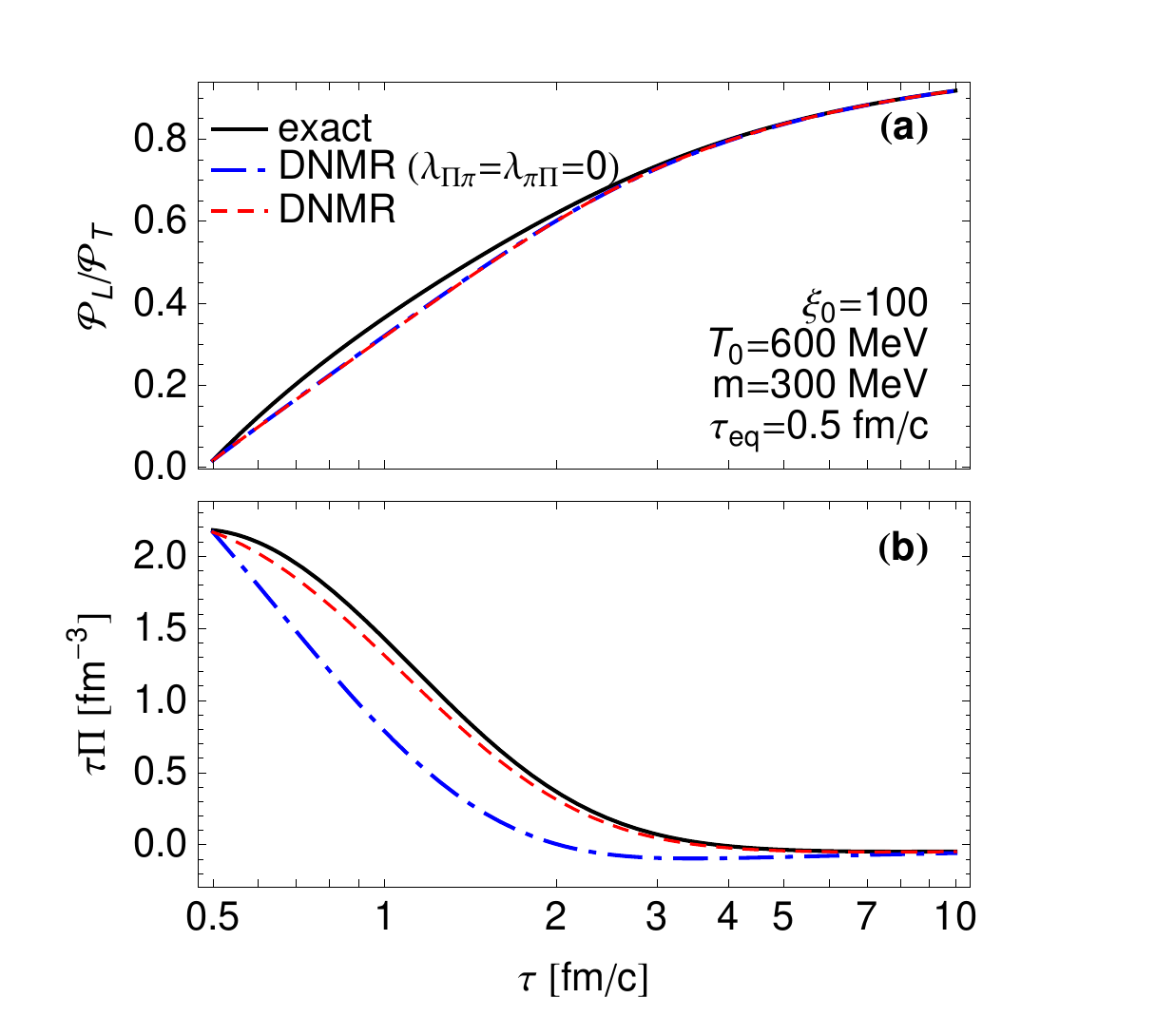}
\hspace{-1.5cm}
\includegraphics[angle=0,width=0.575\textwidth]{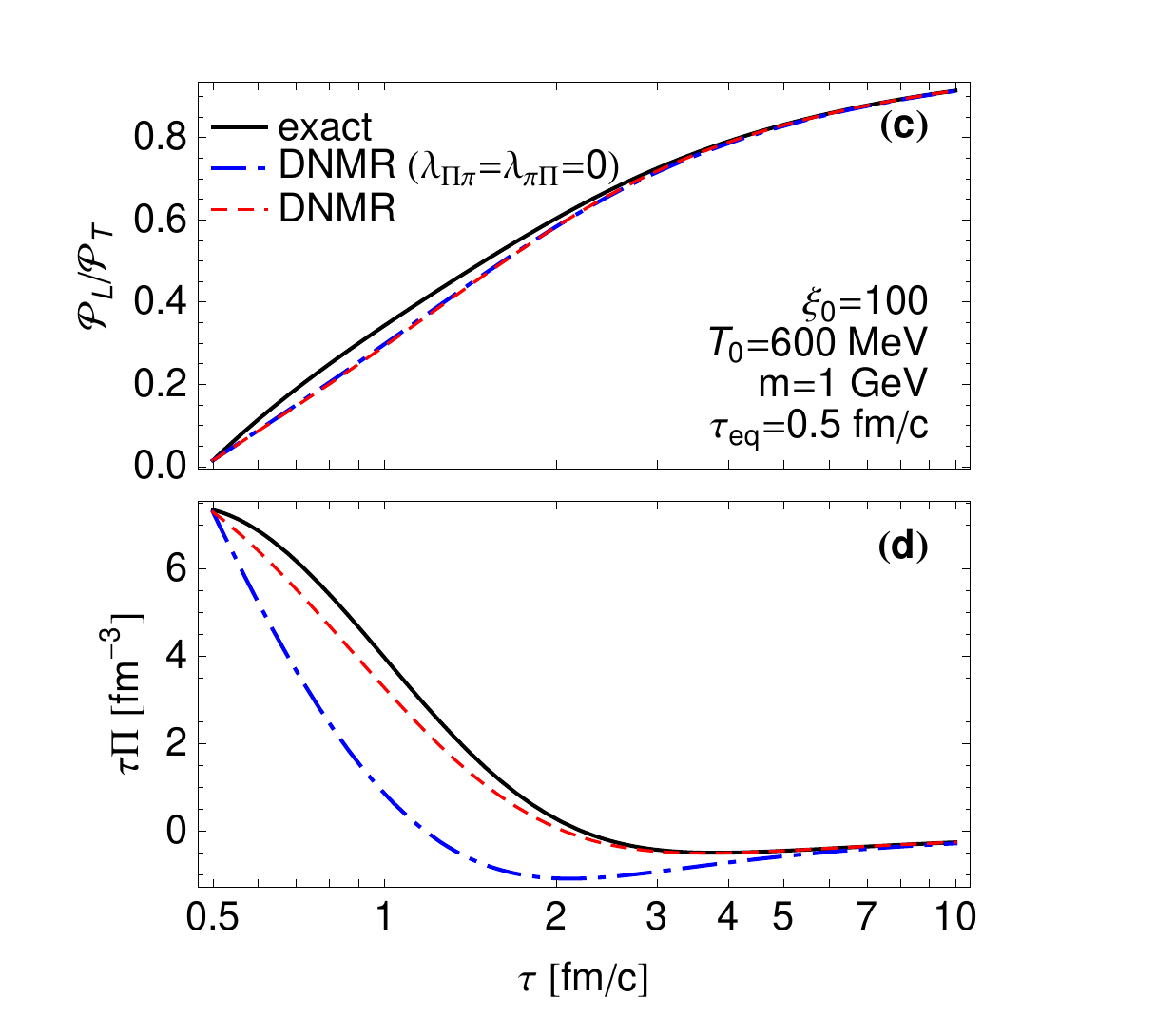}
}
\caption{(Color online) Same as Fig.~\protect\ref{fig:coupl2} except here we
take $\protect\xi_0=100$.}
\label{fig:coupl4}
\end{figure}

In Figs.~\ref{fig:coupl1} -- \ref{fig:coupl4} we compare the proper-time
evolution of the pressure anisotropy $\mathcal{P}_L/\mathcal{P}_T$ (top
panels) and the bulk pressure $\Pi$ multiplied by proper-time $\tau$ (bottom
panels) obtained from the exact solution of the Boltzmann equation (black
solid line), the full second-order viscous equations including the
shear-bulk couplings $\lambda_{\Pi\pi}$ and $\lambda_{\pi\Pi}$ (red dashed
line), and second-order viscous equations with $\lambda_{\Pi\pi}=\lambda_{%
\pi\Pi}=0$ (blue dot-dashed line). We show the results for two different
initial values of the anisotropy parameter $\xi_0 \in \{0,100\}$ and three
values of the particle mass $m\in\{0.03,0.3,1\}$ GeV.

As one can see from Figs.~\ref{fig:coupl1} and \ref{fig:coupl3}, which both
assume $m=30$ MeV, the 14-moment second-order viscous hydrodynamical result
(DNMR)
including the shear-bulk couplings works quite well in reproducing the exact
solution for small masses. For the larger masses shown in Figs.~\ref%
{fig:coupl2} and \ref{fig:coupl4} ($m=300$ MeV and $m=1$ GeV), we see
somewhat larger deviations from the exact solution. Note importantly that in
all cases shown, when one turns off the shear-bulk couplings by setting $%
\lambda _{\Pi \pi }=\lambda _{\pi \Pi }=0$, the resulting bulk pressure
evolution does not agree well with the exact solution demonstrating the
importance of these couplings for early time dynamics. Additionally, one
notices that inclusion of these couplings has a larger relative effect on
the bulk pressure evolution than the pressure anisotropy with the effect on the
pressure anisotropy increasing as the mass increases.


\subsection{Comparison with anisotropic hydrodynamics}

\label{subsect:ahydrocompare} 

\begin{figure}[t]
\centerline{\includegraphics[angle=0,width=0.65\textwidth]{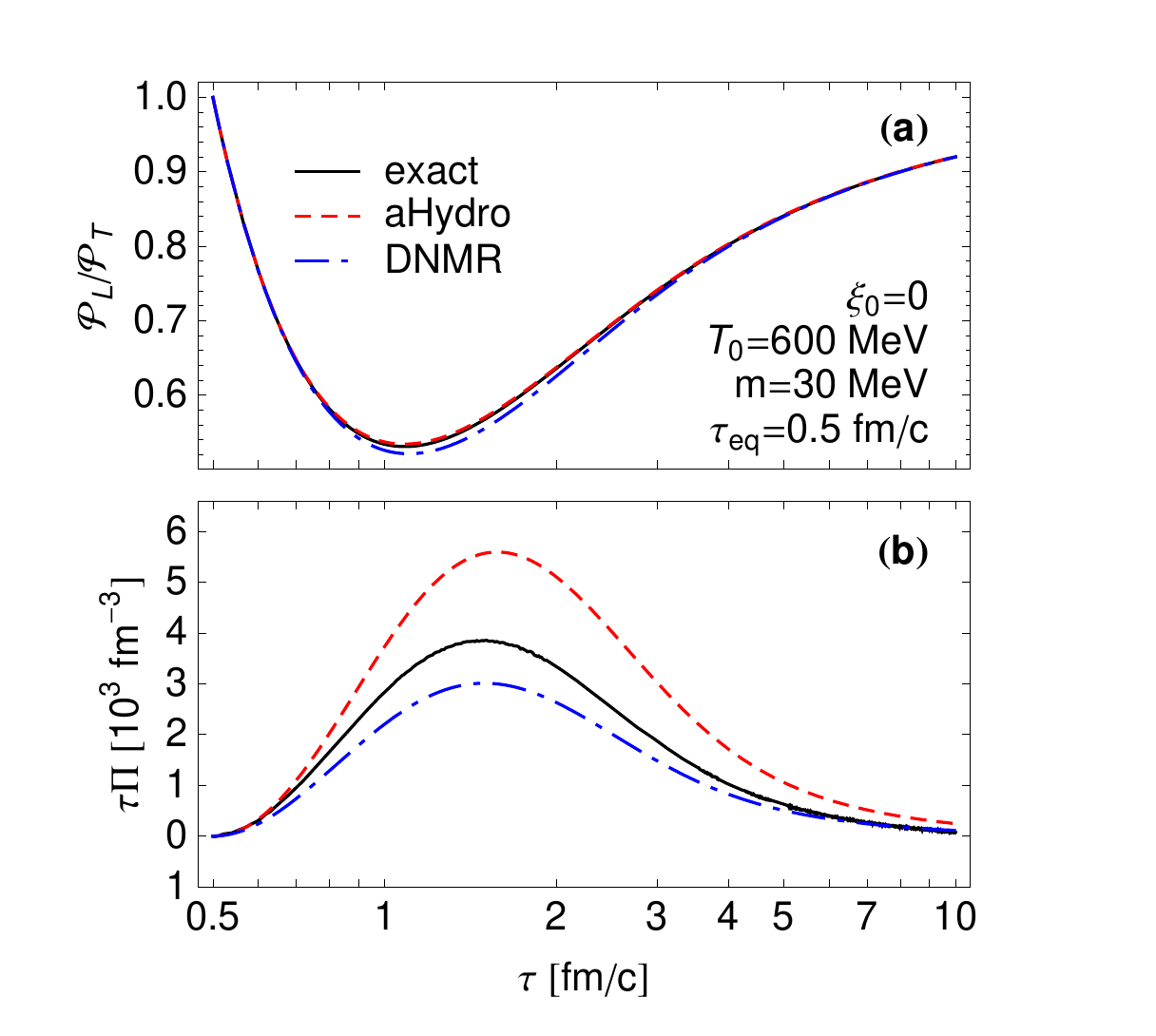}}
\caption{(Color online) Proper-time evolution of pressure anisotropy $%
\mathcal{P}_L/\mathcal{P}_T$ (a) and bulk pressure (b). The three
lines correspond to the exact solution of the Boltzmann equation 
\protect\cite{Florkowski:2014sfa} (black solid line), the full second-order
viscous equations including the shear-bulk
couplings $\protect\lambda_{\Pi\protect\pi}$ and $\protect\lambda_{\protect%
\pi\Pi}$ \protect\cite{Denicol:2014vaa} (blue dot-dashed line), and anisotropic 
hydrodynamics \protect\cite{Nopoush:2014pfa} (red dashed
line). For both Figures we used $m=$ 30 MeV, $\protect\tau_0$ = 0.5 fm/c, $%
\protect\tau_{\mathrm{eq}} =\protect\tau_{\protect\pi} =\protect\tau_{\Pi}$
= 0.5 fm/c, and $T_0$ = 600 MeV. The initial spheroidal anisotropy parameter
for initial distribution function of the exact solution of Boltzmann
equation is taken to be $\protect\xi_0=0$, in consequence $\protect\pi_0=0$
and $\Pi_0$ = 0.}
\label{fig:comp1}
\end{figure}

\begin{figure}[t]
\centerline{
\hspace{1cm}
\includegraphics[angle=0,width=0.575\textwidth]{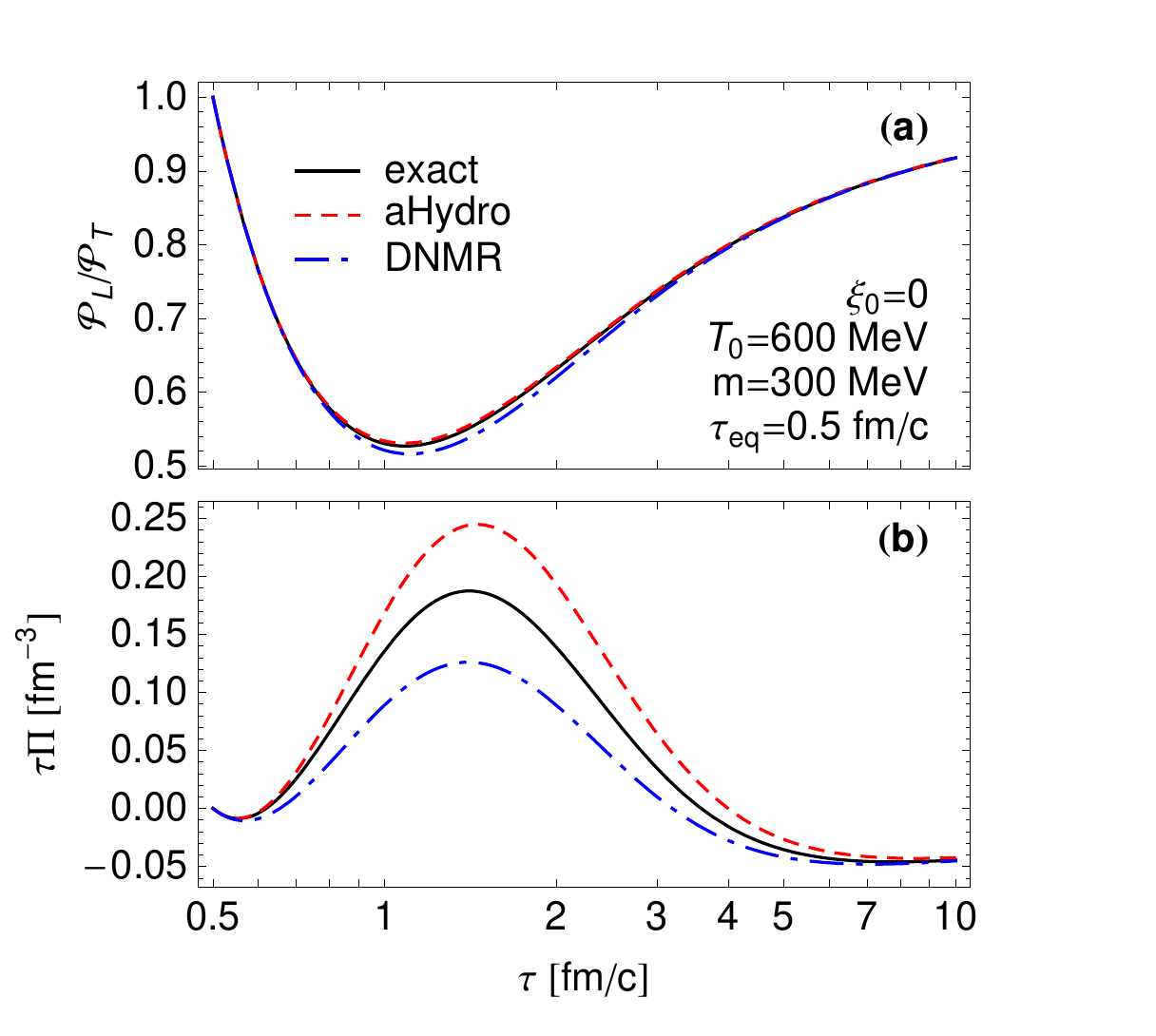}
\hspace{-1.5cm}
\includegraphics[angle=0,width=0.575\textwidth]{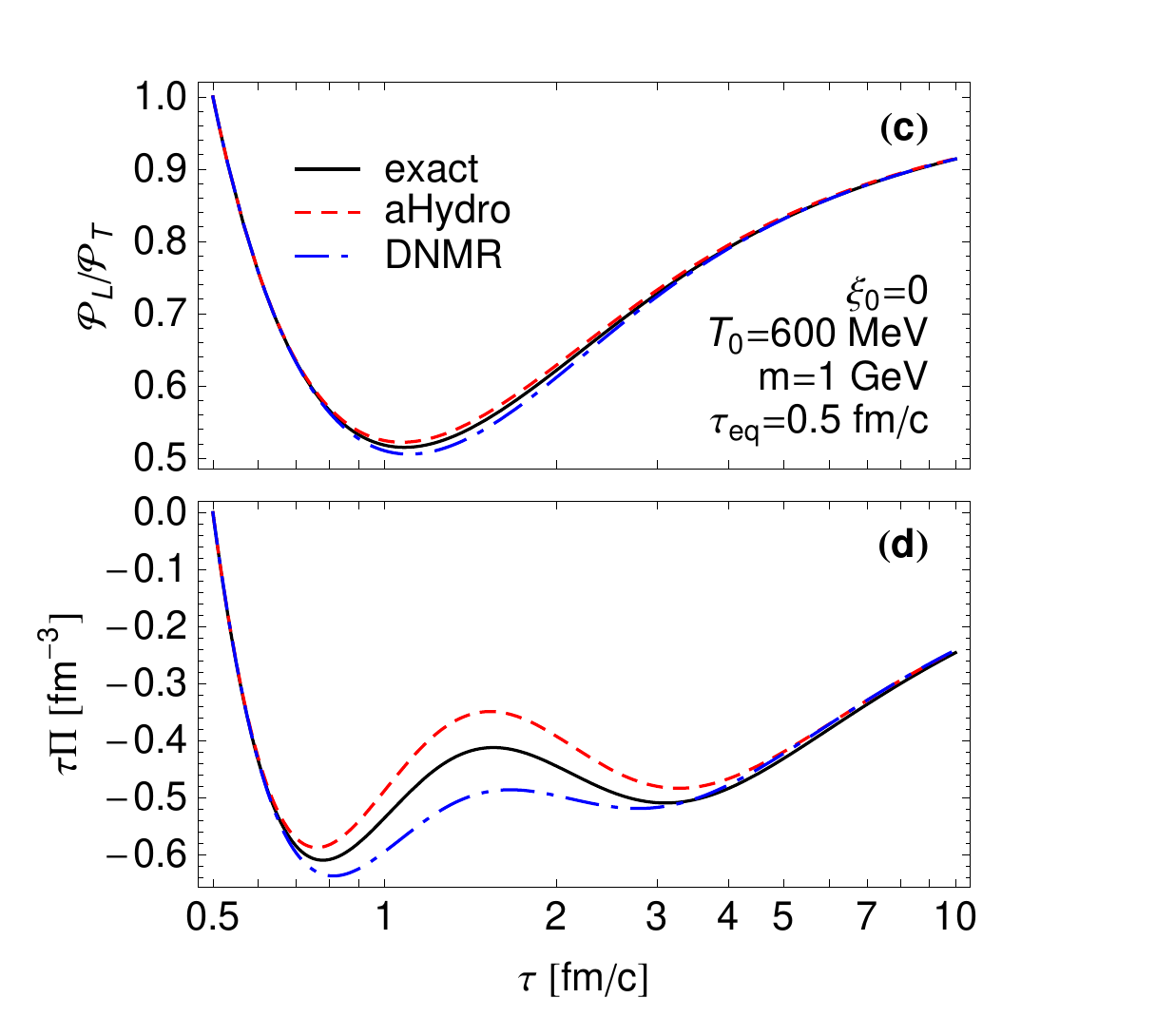}
}
\caption{(Color online) Proper-time evolution of $\mathcal{P}_L/\mathcal{P}%
_T $ ((a) and (c)) and bulk pressure ((b) and (d)). Parameters and descriptions are the
same as in Fig.~\protect\ref{fig:comp1} except here we take $m=300$ MeV
((a) and (b)) and $m=1$ GeV ((c) and (d)).}
\label{fig:comp2}
\end{figure}

In this Section we compare the results of the second-order viscous
hydrodynamics and anisotropic hydrodynamics with the exact solutions of the
RTA Boltzmann equation. In the framework of anisotropic hydrodynamics the
system is characterized by a set of non-equilibrium parameters and one does
not deal explicitly with the kinetic coefficients. Interestingly, one may
demonstrate that both the second-order viscous hydrodynamics and anisotropic
hydrodynamics lead to similar description of the system and the two
approaches agree reasonably well with the exact kinetic solution.

Working within the anisotropic hydrodynamics framework, we numerically solve
Eqs.~(\ref{eq:final0m})--(\ref{eq:final2m}) for the non-equilibrium
parameters $\alpha_x$, $\alpha_z$ and $\lambda$. We fix the initial
conditions for $\alpha_x$, $\alpha_z$, and $\lambda$ such that the initial
energy density, pressure anisotropy, and bulk pressure are the same as those
used in the exact solution and the second-order viscous hydrodynamics
solution. At each step of the numerical integration we use Eq.~(\ref{eq:dlm}%
) to self-consistently determine the effective temperature $T$ which appears
in the equations of motion.

\begin{figure}[t]
\centerline{\includegraphics[angle=0,width=0.65\textwidth]{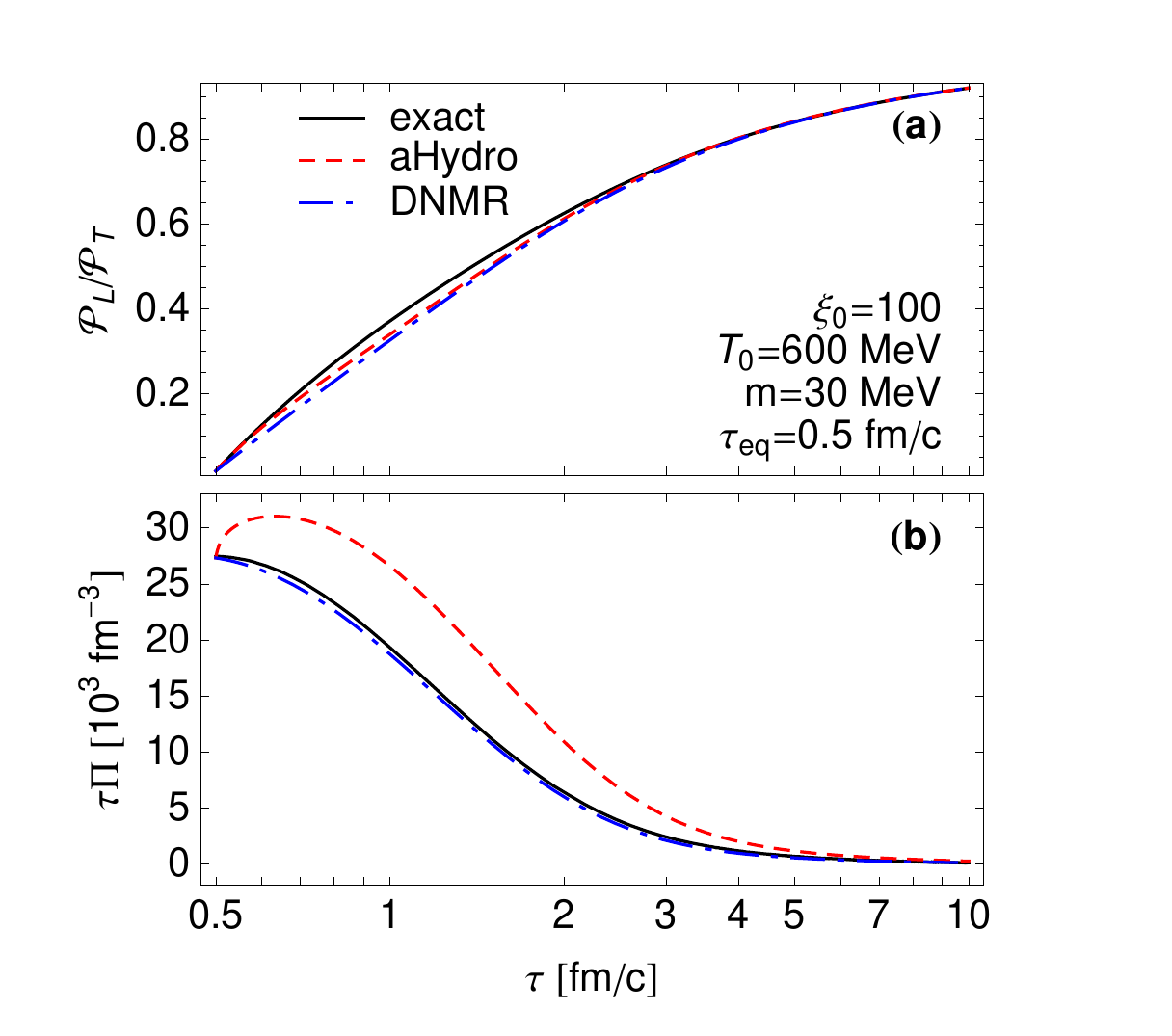}}
\caption{(Color online) Proper-time evolution of $\mathcal{P}_L/\mathcal{P}%
_T $ (a) and bulk pressure (b). Parameters and descriptions are the
same as in Fig.~\protect\ref{fig:comp1} except here we take $\protect\xi%
_0=100$.}
\label{fig:comp3}
\end{figure}

\begin{figure}[t]
\centerline{
\hspace{1cm}
\includegraphics[angle=0,width=0.65\textwidth]{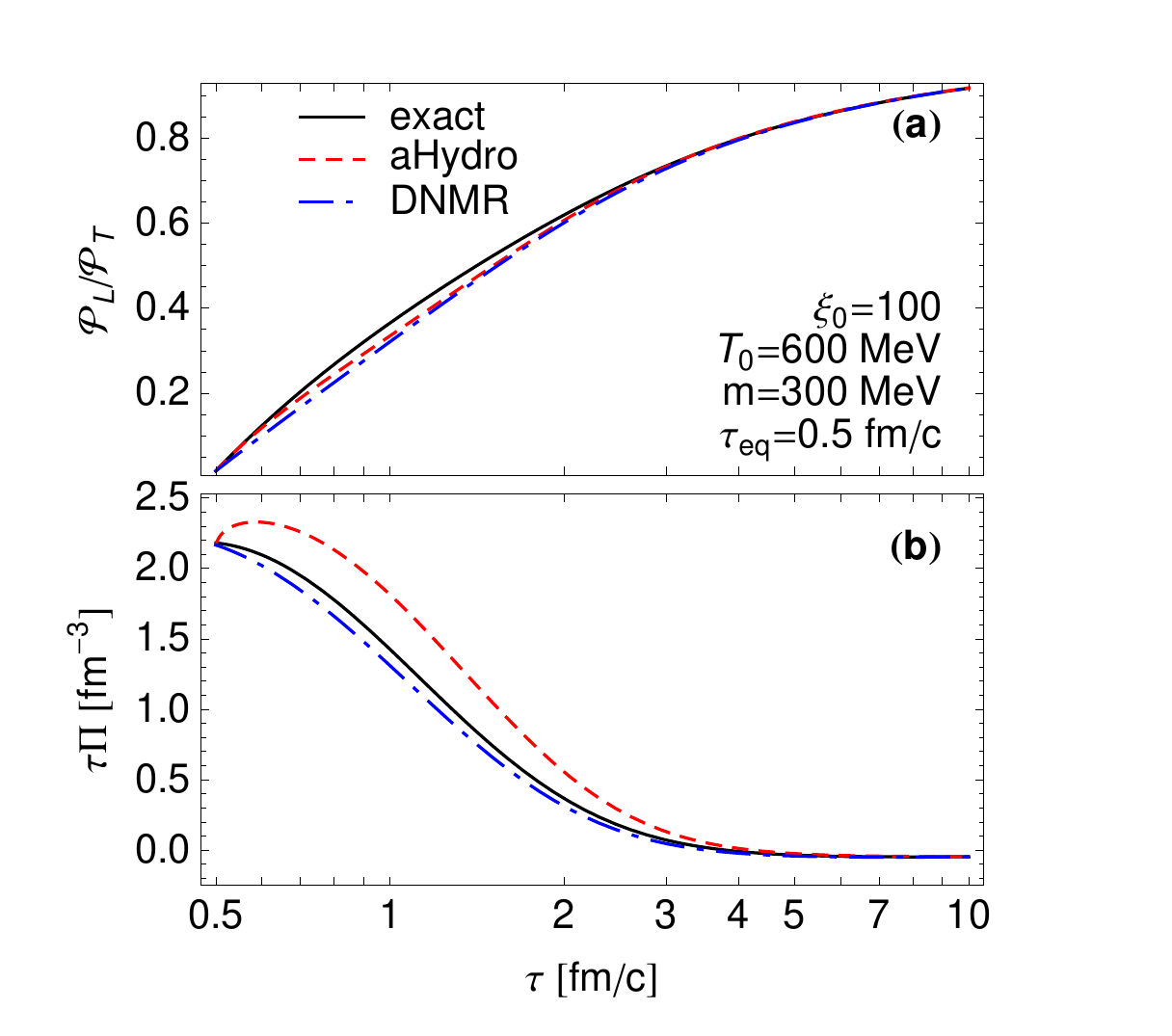}
\hspace{-1.5cm}
\includegraphics[angle=0,width=0.65\textwidth]{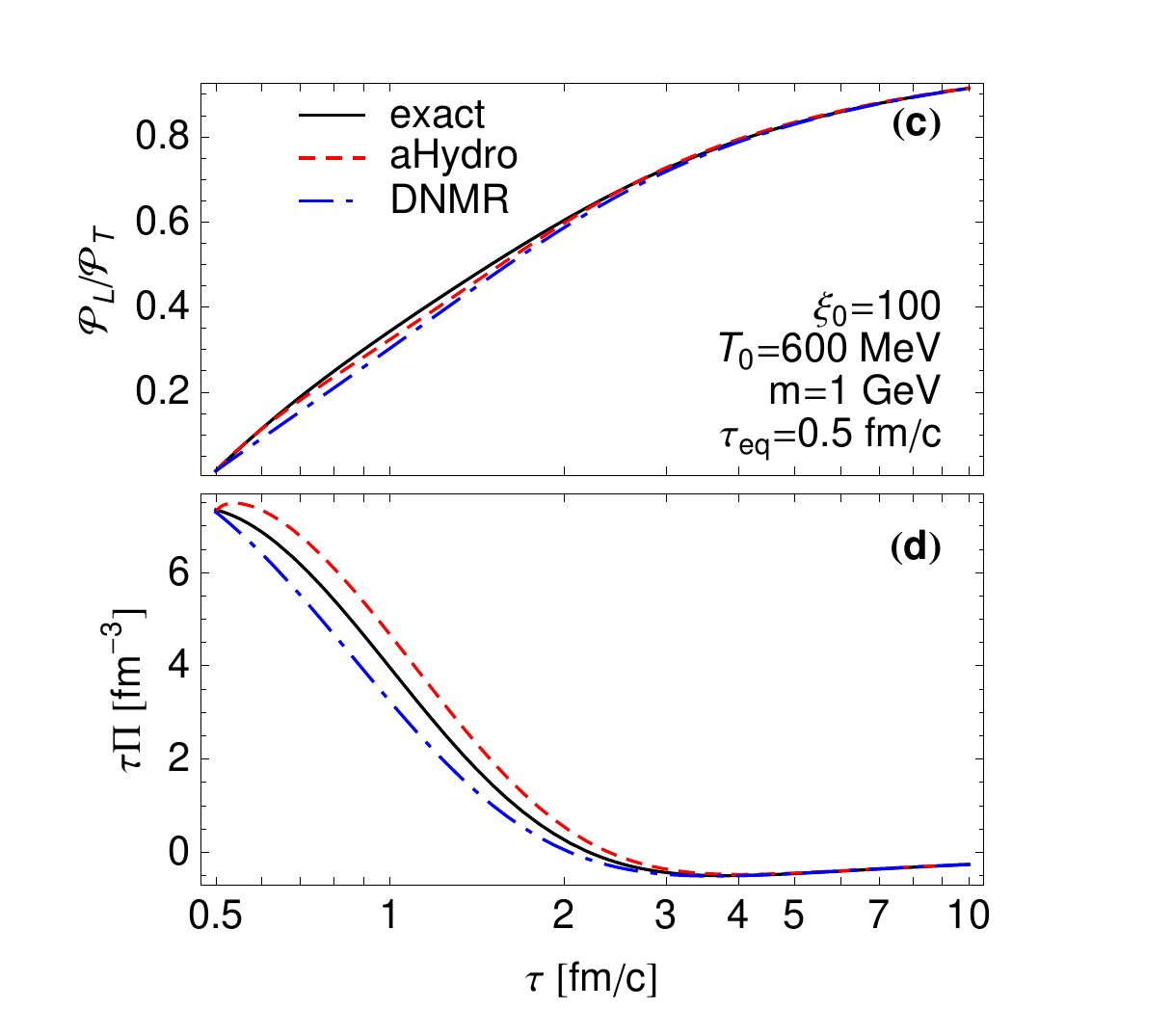}}
\caption{(Color online) Proper-time evolution of $\mathcal{P}_L/\mathcal{P}%
_T $ ((a) and (c)) and bulk pressure ((b) and (d)). Parameters and descriptions are the
same as in Fig.~\protect\ref{fig:comp2} except here we take $\protect\xi%
_0=100$.}
\label{fig:comp4}
\end{figure}

Our comparisons between second-order viscous hydrodynamics and anisotropic
hydrodynamics are presented in Figs.~\ref{fig:comp1} -- \ref{fig:comp4}. The
parameters are chosen to be the same as in the previous Section. From these
figures, one sees that anisotropic hydrodynamics provides a comparable
description of bulk and shear pressure as complete second-order viscous
hydrodynamics. However, in the small mass case it seems that second-order
viscous hydrodynamics does a better job in reproducing the evolution of the
bulk pressure for large initial anisotropies. In most cases, however
anisotropic hydro does a slightly better job in reproducing the exact
solution for the pressure anisotropy. Note, however, that herein we have
assumed $\tau _{\mathrm{eq}}$ = 0.5 fm/c in all figures. If one were to take
larger values of $\tau _{\mathrm{eq}}$ or smaller initial temperatures, then
one would have to reconsider this comparison.



\section{Conclusions and Outlook}

\label{sect:conc} 

In this paper we have demonstrated the importance of shear-bulk coupling in
the early time dynamics of the quark gluon plasma. These couplings are
important because there are large shear corrections at early times and these
seem to have a marked effect on the evolution of the bulk viscous pressure.
The reverse effect of bulk pressure on the shear pressure, measured here in
terms of the pressure anisotropy, was found to be small. To reach this
conclusion, we compared the results of second-order viscous hydrodynamics
using a complete 14-moment approximation with exact solutions to the 0+1d
kinetic equations in relaxation time approximation. We found that without
the shear-bulk coupling one is not able to reproduce the behavior exhibited
by the exact solution.

Following this, we then compared the resulting full 14-moment second-order
viscous hydrodynamics results with recently obtained anisotropic
hydrodynamics evolution equations which include a bulk degree of freedom. We
demonstrated that both the complete second-order viscous hydrodynamics
framework and anisotropic hydrodynamics were able to reproduce the exact
result with comparable accuracy. For small masses, the 14-moment approximation
has better agreement with the bulk pressure evolution than anisotropic
hydrodynamics; however, anisotropic hydrodynamics was found to better
reproduce the pressure anisotropy in this case. For larger masses, both
approaches had comparable accuracy.

Looking forward, herein we showed explicitly that shear-bulk couplings can
be important for the early time dynamics of the bulk pressure in simulations
of relativistic heavy ion collisions. It will be interesting to extend the
results contained herein to higher dimensional systems in order to gauge the
full impact that shear-bulk couplings have on the dynamical evolution of the
system. In the case of anisotropic hydrodynamics, the full 1+1d equations
including the effects of the bulk pressure have already appeared in the
literature \cite{Nopoush:2014pfa}. For 14-moment second-order viscous
hydrodynamics, the general equations are known even for 3+1d including bulk
viscous effects \cite{Denicol:2014vaa}. It will be interesting to see what
the impact shear-bulk couplings will be in both cases. We leave this for
future work.

\acknowledgments{
We thank A. Jaiswal for very useful discussions.
G.S.D. was supported by a Banting Fellowship from the Natural Sciences and Engineering Research Council of Canada.
W.F. was supported by Polish National Science Center grant No. DEC-2012/06/A/ST2/00390.
R.R. was supported by Polish National Science Center grant No.~DEC-2012/07/D/ST2/02125 and U.S.~DOE Grant No.~DE-SC0004104.   M.S. was supported in part by U.S.~DOE Grant No.~DE-SC0004104.
}

\appendix


\section{Thermodynamic integrals}

\label{sect:integrals} 

The integrals defined in Eq.~(\ref{intI}) can be written in the following
form 
\begin{eqnarray}
I_{n q}(T,m) &=&\frac{N_{\mathrm{dof}}}{2 \pi ^2 (2 q+1)\text{!!}}
\int_{m}^{\infty}e^{-\frac{t}{T}} \left(t^2-m^2\right)^{\frac{1}{2} (2 q+1)}
t^{n-2 q}\,dt \, ,
\end{eqnarray}
which leads to the following results 
\begin{eqnarray}
I_{-2,0}(T,m) &=& 4\pi\tilde{N} \left[K_0(\hat{m}_{\mathrm{eq}})-\hat{m}_{%
\mathrm{eq}} \left(K_1(\hat{m}_{\mathrm{eq}})-K_{i1}(\hat{m}_{\mathrm{eq}%
})\right) \right], \\
I_{-1,0}(T,m) &=& 4\pi\tilde{N} T \hat{m}_{\mathrm{eq}} \left[K_1(\hat{m}_{%
\mathrm{eq}})-K_{i1}(\hat{m}_{\mathrm{eq}})\right], \\
I_{0,0}(T,m) &=& 4\pi\tilde{N} T^2 \hat{m}_{\mathrm{eq}} K_1(\hat{m}_{%
\mathrm{eq}} ) \, , \\
I_{1,0}(T,m) &=& 4\pi\tilde{N} T^3 \hat{m}_{\mathrm{eq}} ^2 K_2(\hat{m}_{%
\mathrm{eq}} ) \, , \\
I_{2,0}(T,m) &=& 4\pi\tilde{N} T^4 \hat{m}_{\mathrm{eq}} ^2 (\hat{m}_{%
\mathrm{eq}} K_1(\hat{m}_{\mathrm{eq}} )+3 K_2(\hat{m}_{\mathrm{eq}} )) \, ,
\\
I_{2,1}(T,m) &=& 4\pi\tilde{N} T^4 \hat{m}_{\mathrm{eq}} ^2 K_2(\hat{m}_{%
\mathrm{eq}} ) \, , \\
I_{2,2}(T,m) &=& 4\pi\tilde{N} \, \frac{T^4 \hat{m}_{\mathrm{eq}}^2}{30} \! %
\Big[\!\left(6-m_{\mathrm{eq}}^2\right) K_2(m_{\mathrm{eq}}) 
\nonumber \\
&& \left. \hspace{1.85cm} + m_{\mathrm{eq}}^2 \left( 3 K_0(m_{\mathrm{eq}}) - 2 m_{\mathrm{eq}} \Big( K_1(m_{\mathrm{eq%
}}) - K_{i,1}(m_{\mathrm{eq}}) \right) \Big) \right] ,  \\
I_{3,0}(T,m) &=& 4\pi\tilde{N} T^5 \hat{m}_{\mathrm{eq}} \left(\hat{m}_{%
\mathrm{eq}} \left(\hat{m}_{\mathrm{eq}} ^2+12\right) K_0(\hat{m}_{\mathrm{eq%
}} )+\left(5 \hat{m}_{\mathrm{eq}} ^2+24\right) K_1(\hat{m}_{\mathrm{eq}}
)\right) \, , \\
I_{3,1}(T,m) &=& 4\pi\tilde{N} T^5 \hat{m}_{\mathrm{eq}} ^3 K_3(\hat{m}_{%
\mathrm{eq}} ) \, , \\
I_{4,1}(T,m) &=& 4\pi\tilde{N} T^6 \hat{m}_{\mathrm{eq}} \left(\hat{m}_{%
\mathrm{eq}} \left(\hat{m}_{\mathrm{eq}} ^2+20\right) K_0(\hat{m}_{\mathrm{eq%
}} )+\left(7 \hat{m}_{\mathrm{eq}} ^2+40\right) K_1(\hat{m}_{\mathrm{eq}}
)\right) \, , \\
I_{4,2}(T,m) &=& 4\pi\tilde{N} T^6 \hat{m}_{\mathrm{eq}} ^3 K_3(\hat{m}_{%
\mathrm{eq}} ) \, ,
\end{eqnarray}
where $I_{2,1} = \mathcal{P}$, $I_{2,0} = \mathcal{E}$, and $I_{3,0} = T^2
(\partial \mathcal{E}/\partial T)$. The function $K_{i,1}(z)$ is defined by
the integral 
\begin{equation}
K_{i,1}(z)= \int_0^{\infty} \frac{\mathrm{e}^{- z \cosh t}}{\cosh t} \, dt
\, ,  \label{Kin}
\end{equation}
and can be expressed as \cite{Florkowski:2014sfa} 
\begin{equation}
K_{i,1}(z) = \frac{\pi}{2} \left[1 - z K_0(z) L_{-1}(z) - z K_1(z) L_{0}(z) %
\right] ,  \label{Kinan}
\end{equation}
where $L_i$ is a modified Struve function.

\bibliography{shearbulk}

\end{document}